\documentclass[10pt,twocolumn,twoside]{IEEEtran}

\usepackage{multirow}

\usepackage{times}
\usepackage{amsmath,dsfont}
\usepackage{amssymb,amsthm}
\usepackage{epsfig,verbatim}
\usepackage{subfigure}
\usepackage{setspace}
\usepackage{color}
\usepackage{cite}
\usepackage{epstopdf}
\usepackage{graphics}
\usepackage{accents}
\usepackage{acronym}
\usepackage[bookmarks,colorlinks]{hyperref}
\usepackage{booktabs}
\usepackage{mathtools}
\usepackage{algorithm}
\usepackage{enumitem}
\usepackage{balance}

\usepackage{algorithmicx}
\usepackage[noend]{algpseudocode}

\allowdisplaybreaks[4]

\newcommand{\mySet}[1]{\mathcal{#1}}
\newcommand{\E}{\mathds{E}}		 			% Stochastic expectation
\newtheorem{theorem}{Theorem}

\newtheorem{lemma}{Lemma}

\definecolor{NewColor}{rgb}{0,0,0} %{0.2,0,0.5}

\acrodef{adc}[ADC]{analog-to-digital convertor}
\acrodef{cs}[CS]{compressed sensing}
\acrodef{dtft}[DTFT]{discrete-time Fourier transform}
\acrodef{dnn}[DNN]{deep neural network}
\acrodef{ml}[ML]{machine learning}
\acrodef{mmse}[MMSE]{minimum \ac{mse}}
\acrodef{csi}[CSI]{channel state information}
\acrodef{map}[MAP]{maximum a-posteriori probability}
\acrodef{snr}[SNR]{signal-to-noise ratio}
\acrodef{bs}[BS]{base station} 
\acrodef{mimo}[MIMO]{multiple-input multiple-output}
\acrodef{mse}[MSE]{mean-squared error}
\acrodef{pdf}[PDF]{probability density function}
\acrodef{rv}[RV]{random variable}
\acrodef{lti}[LTI]{linear time-invariant}
\acrodef{wss}[WSS]{wide-sense stationary}
\acrodef{psd}[PSD]{power spectral density}
\acrodef{ser}[SER]{symbol error rate} 
\acrodef{isi}[ISI]{intersymbol interference} 
\acrodef{lstm}[LSTM]{long short-term memory} 
\acrodef{em}[EM]{expectation minimization} 
\acrodef{tdd}[TDD]{time division duplexing} 
\acrodef{ut}[UT]{user terminal} 
\acrodef{awgn}[AWGN]{additive white Gaussian noise}
\acrodef{cgac}[CGAC]{Complex-gain analog combiner}
\acrodef{psoac}[PSOAC]{Phase-shifter-only analog combiner}
\acrodef{fpga}[FPGA]{field-programmable gate array}
\acrodef{gui}[GUI]{graphical user interface}
\acrodef{magiq}[MaGiQ]{minimal gap iterative quantization}

\IEEEoverridecommandlockouts

\begin{document}

\title{RF Chain Reduction for MIMO Systems: A Hardware Prototype}

\author{Tierui Gong,~\IEEEmembership{Student~Member,~IEEE},
	Nir Shlezinger,~\IEEEmembership{Member,~IEEE},\\
    Shahar Stein Ioushua,~\IEEEmembership{Student~Member,~IEEE},
	Moshe Namer,\\
	Zhijia Yang,
	and Yonina C. Eldar,~\IEEEmembership{Fellow,~IEEE}
\thanks{This work was supported in part by the National Key Research and Development Program of China under grant 2017YFA0700304, by the Israel Science Foundation under grant 0100101, and by the European Union's Horizon 2020 research and innovation program under grant no. 646804-ERC-COG-BNYQ. %\textit{(Corresponding Authors: Nir Shlezinger and Zhijia Yang)}
			  }
\thanks{T. Gong and Z. Yang are with the Key Laboratory of Networked Control Systems, State Key Laboratory of Robotics, Shenyang Institute of Automation, Chinese Academy of Sciences, Shenyang 110016, China, and Institutes for Robotics and Intelligent Manufacturing, Chinese Academy of Sciences, Shenyang 110169, China. T. Gong is also with the University of Chinese Academy of Sciences, Beijing 100049, China (email: gongtierui@sia.cn; yang@sia.ac.cn).
			  }% <-this % stops a space
\thanks{N. Shlezinger  and Y. C. Eldar are with the Faculty of Mathematics and Computer Science, Weizmann Institute of Science, Rehovot, Israel (e-mail: nirshlezinger1@gmail.com; yonina@weizmann.ac.il).
			  }
\thanks{S. S. Ioushua is with the School of Electrical Engineering, Tel Aviv University, Tel Aviv, Israel (e-mail: shunbunstein@gmail.com).
			  }
\thanks{M. Namer is with the Faculty of Electrical Engineering, Technion, Haifa, Israel (e-mail: namer@ee.technion.ac.il).
			  }
			  \vspace{-0.5cm}
}

%\markboth{IEEE Systems Journal,~Vol.~XX, No.~XX, XXX~2020}%
%{Gong \MakeLowercase{\textit{et al.}}: RF Chain Reduction for MIMO Systems: A Hardware Prototype}

\maketitle

\begin{abstract}
	Radio frequency (RF) chain circuits play a major role in digital receiver architectures, allowing passband communication signals to be processed in baseband. When operating at high frequencies, these circuits tend to be costly.  This increased cost imposes a major limitation on future multiple-input multiple-output (MIMO) communication technologies. A common approach to mitigate the increased cost is to utilize hybrid architectures, in which the received signal is combined in analog into a lower dimension, thus reducing the number of RF chains. In this work we study the design and hardware implementation of hybrid architectures via minimizing channel estimation error. We first derive the optimal solution for complex-gain combiners and propose an alternating optimization algorithm for phase-shifter combiners. We then present a hardware prototype implementing analog combining for RF chain reduction. The prototype consists of a specially designed configurable combining board as well as a dedicated experimental setup. Our hardware prototype allows evaluating the effect of analog combining in MIMO systems using actual communication signals. The experimental study, which focuses on channel estimation accuracy in MIMO channels, demonstrates that using the proposed prototype, the achievable channel estimation performance is within a small gap  in a statistical sense from that obtained using a costly receiver in which each antenna is connected to a dedicated RF chain. %Furthermore, in the considered scenarios, this gap becomes negligible when the reduction rate, i.e., the ratio of the number of RF chains to the number of antennas, is above $62.5\%$.
\end{abstract}


\IEEEpeerreviewmaketitle

%----------------------------------------------------------------------------------------
%	Introduction
%----------------------------------------------------------------------------------------
\section{Introduction}
	Next generation wireless systems are required to meet a growing throughput demand \cite{andrews2014will}. Two key technologies offer the potential of providing dramatic increase in spectral efficiency. The first scheme equips the cellular \acp{bs} with a large number of antennas, realizing a massive \ac{mimo} system \cite{marzetta2010noncooperative, shlezinger2018spectral, Gong2019Compressive}. The second method explores the millimeter wave frequency band, overcoming the congestion of the conventional spectrum \cite{xiao2017millimeter}. 
	A major drawback of utilizing these technologies stems from the fact that radio frequency (RF) chain circuits, which allow the passband communication signals to be processed in baseband and are thus an essential component in digital receiver architectures, lead to increased cost and power consumption  when operating at high carrier frequencies. This increased cost becomes a major practical bottleneck when implementing \ac{mimo} antenna arrays operating at millimeter wave bands in which each antenna is connected to an RF chain.
	
One of the common approaches to mitigate this increased cost is to utilize fewer RF chains \cite{Zhang2005Variable,Ayach2014Spatially,Alkhateeb2014Channel}, namely, implementing RF chain reduction. This reduction is carried out by introducing additional analog hardware, which most commonly consists of controllable phase shifters and adders \cite{Zhang2005Variable}, typically designed to optimize the achievable rate \cite{Ayach2014Spatially} or the channel estimation accuracy \cite{Alkhateeb2014Channel}.
	In such systems, the analog signal observed at the antenna array is combined into a lower dimension digital signal using dedicated hardware \cite{Méndez2016Hybrid}. When utilizing such hybrid architectures, the number of RF chains, and hence the number of inputs processed in the digital domain, is smaller than the number of antennas. 
	Receiver designs implementing RF chain reduction using analog combining  have been the focus of extensive research in recent years \cite{choi2017resolution,Méndez2016Hybrid,shlezinger2018asymptotic,shlezinger2019dynamic, Ioushua2019Hybrid, Ioushua2018Pilot,alkhateeb2014mimo, mo2017hybrid, roth2018comparison, chai2018antenna, abbas2017millimeter, lee2015hybrid,kim2015mse,shlezinger2018hardware, karamalis2006adaptive}. These previous works include the design of different analog combining structures \cite{Méndez2016Hybrid,Ioushua2019Hybrid, Ioushua2018Pilot,alkhateeb2014mimo, lee2015hybrid,kim2015mse}, hybrid architectures with low-resolution analog-to-digital conversion \cite{shlezinger2018asymptotic,choi2017resolution, mo2017hybrid, roth2018comparison, chai2018antenna, abbas2017millimeter,shlezinger2018hardware}, and the integration of dynamic analog combining as part of the physical antenna technology \cite{shlezinger2019dynamic}.
	The theoretical maturity of the concept of \ac{mimo} communications with RF chain reduction suggests the need to demonstrate and evaluate the implementation of such systems in hardware, which is the focus of this work.
	
	We present a prototype of a configurable analog combining hardware board. Our analog combiner hardware can realize different RF chain reduction strategies, including complex gain combiners \cite{karamalis2006adaptive, shlezinger2018hardware}, phase shifter networks \cite{Ioushua2019Hybrid, Méndez2016Hybrid}, and antenna selection techniques \cite{chai2018antenna}. The board is designed to experimentally evaluate \ac{mimo} communications with RF chain reduction applied to actual up-converted passband signals. 
	Similarly to \cite{kuehne2018analog,mondal201825}, we design a dedicated analog combiner hardware for experimental purposes.  While \cite{mondal201825} used RF integrated circuits, focusing on millimeter wave bands, our prototype targets the  sub-6 GHz frequency range, as in \cite{kuehne2018analog}. The main advantage of our prototype over \cite{kuehne2018analog} is in the improved phase resolution, which stems from the usage of dedicated vector multipliers, allowing to more accurately set the phase of the complex coefficients and thus to better test analytically derived analog combiners. While our prototype supports a smaller number of antennas compared to \cite{kuehne2018analog},  we  propose a method for using it to experiment in scenarios with larger number of antennas using a virtual channel extension.   
  
	Our main design criteria is  the channel estimation accuracy, using an analog combining configuration algorithm which extends and improves upon the method suggested in \cite{Ioushua2019Hybrid}. 
	The proposed algorithm uses alternating optimization to obtain a suitable analog combining configuration under structure constraints, such as the requirement to utilize unit gain coefficients arising in phase shifter networks. 
	Our method achieved closer feasible approximations of the optimal unconstrained combiner compared to the algorithm of \cite{Ioushua2019Hybrid} by introducing an additional degree of freedom which is exploited to improve the approximation under structure constraints.
	While we focus on the task of channel estimation, the proposed prototype can be used to implement analog combiners designed according to alternative objectives, e.g., maximize the achievable rate as in \cite{mo2017hybrid}, or signal recovery, as suggested in \cite{Ioushua2019Hybrid}.

	The proposed hardware prototype, which combines our specially designed analog combiner board with a dedicated experimental hardware setup, demonstrates the feasibility of hybrid architectures in wireless networks. 
	In particular, we show that, using a phase shifter network configured with our analog combiner board and the proposed design algorithm, one can achieve channel estimation accuracy which is within a small gap from that achievable using a \ac{mimo} receiver with a complex controllable gain combiner. %Furthermore, the channel estimation accuracy gap from that achievable using a costly fully-digital \ac{mimo} receiver, in which each antenna is connected to a dedicated RF chain, is shown to become negligible when the RF chain  reduction rate is above $62.5\%$ for the considered scenarios.

	The main contributions of the paper are summarized below:
	\begin{itemize}
	    \item
	    We study the design of analog combiners to maximize channel estimation accuracy. To that aim, we first derive the optimal solution for complex-gain hybrid combiners in the noisy setting, extending the derivation in \cite[Sec. IV]{Ioushua2018Pilot} which considered noiseless settings. We show that the combiner is not unique, and specializes to that of \cite{Ioushua2018Pilot} in the high \ac{snr} regime.

		\item 
		Then, we propose an algorithm for designing constrained hybrid beamformers, focusing on the common phase-shifter network architecture \cite{Méndez2016Hybrid}. Our method is based on alternating optimization, i.e., joint optimization by iteratively optimizing over one quantity while keeping the other variables fixed extending the previously proposed method of \cite{Ioushua2019Hybrid}. Our algorithm builds upon the non-uniqueness of the complex-gain solution to find the one which can be best approximated using a feasible combiner setting.

		\item 
		Finally, we present a dedicated  hardware prototype operating in the sub-6 GHz band, with configurable complex gains of extremely high phase resolution, allowing reliable comparison of different hybrid architectures using passband signals. The prototype is used for evaluating our method, demonstrating its ability to approach the performance of variable gain combiners using constrained phase shifters.
	\end{itemize}

	% Organization paragraph
	The rest of this paper is organized as follows: In Section~\ref{sec:BTP}, we formulate the model for wireless communication with RF chain reduction and present the channel estimation problem. Section~\ref{sec:Algorithm} details the algorithm implemented for configuring the analog combiner, and Section~\ref{sec:SA} presents a detailed description of the prototype and  its components. Experimental results are given in Section~\ref{sec:ER}, and Section~\ref{sec:Conclusions} provides concluding remarks.

	% Notations paragraph
	Throughout the paper, we use boldface lower-case letters to denote vectors, e.g., ${\mathbf{x}}$,
	and the $i$th element of ${\mathbf{x}}$ is written as $\left( {\mathbf{x}}\right) _i$. Boldface upper-case letters are used for matrices, e.g. $\mathbf{M}$, whose $(i,k)$th entry is $(\mathbf{M})_{i,k}$ and $i$th column is $(\mathbf{M})_{:,i}$. 
	We use ${\rm vec}(\cdot)$ to denote the vectorization operator, $\mathbf{I}_n$ is the $n\times n$ identity matrix, $\otimes$ is the Kronecker product, $\| \cdot \|$ is the $l_2$ norm, $\| \cdot \|_F$ is the Frobenius norm, $tr(\cdot)$ is the trace operator, while $(\cdot)^T$ and $(\cdot)^*$ denote the transpose and complex transpose, respectively. The proper-complex Gaussian distribution is denoted as $\mathcal{CN}$, and  
	$\mathbb{C}$ is the set of complex numbers. The expectation of a random variable $x$ is represented by the operator $\E\{x\}$.

%----------------------------------------------------------------------------------------
%	Basic theory of the prototype
%----------------------------------------------------------------------------------------
\section{System Model}
\label{sec:BTP}
	Our hardware prototype implements RF chain reduction for \ac{mimo} receivers. In particular, we design the hybrid architecture to facilitate channel estimation in multi-antenna cellular \acp{bs}. To formulate the setup and the design objectives, we first detail the problem formulation in Subsection \ref{subsec:PF}, after which we present the considered model for the unknown channel in Subsection \ref{subsec:CM}.

	\subsection{Problem Formulation}
	\label{subsec:PF}
	Consider a single-cell network in which a \ac{bs} is equipped with $N_{bs}$ antennas and serves $K$ single-antenna \acp{ut}.  We focus on the uplink, namely, the transmission from the \acp{ut} to the \ac{bs}. The \ac{bs} utilizes an analog combiner, and thus observes the channel output after it has been linearly combined and acquired using $N_{rf} \le N_{bs}$ RF chains. The analog combiner network is denoted via the matrix $\mathbf{W} \in \mathcal{W}$, where $\mathcal{W} \subseteq \mathbb{C}^{N_{rf} \times N_{bs}}$ represents the feasible set of analog combiners. This set is determined by the specific hardware of the analog combiner, and can represent, e.g., complex gains and phase shifter networks \cite{Méndez2016Hybrid, Ioushua2019Hybrid}.
	
\begin{figure*}[t!]
		\centering
		\includegraphics[width=12.5cm]{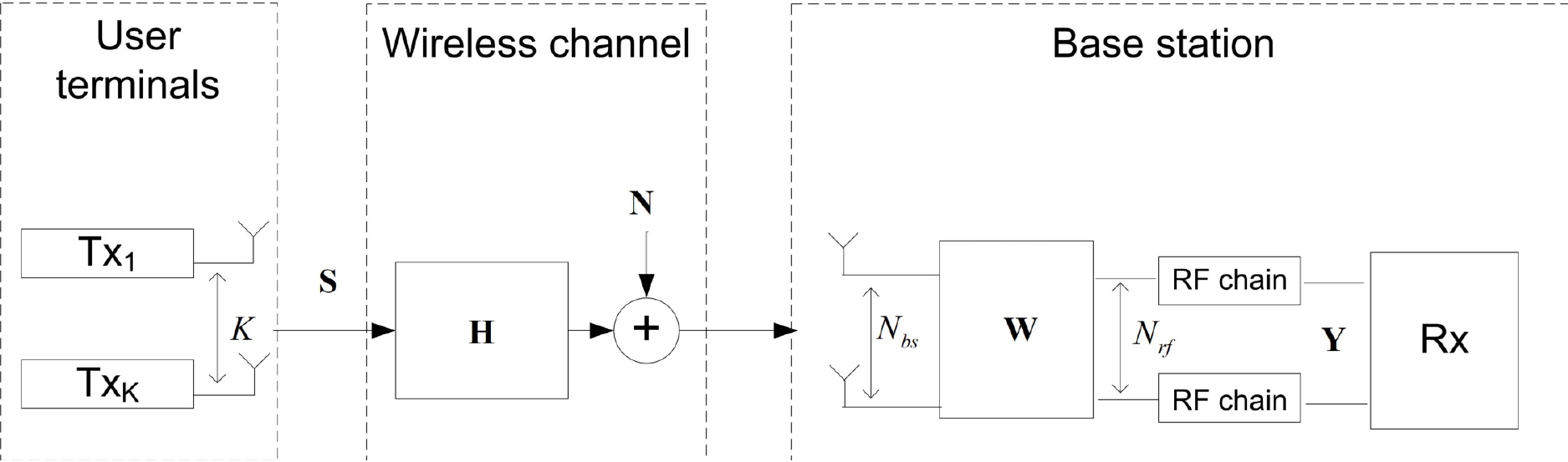}
		\caption{Baseband communication system diagram.}
		\label{ModelPicture}
		%\vspace{-1em}
	\end{figure*}

	Let $\mathbf{H} \in \mathbb{C}^{N_{bs} \times K}$ denote the wireless channel matrix and $\mathbf{S} \in \mathbb{C}^{\tau \times K}$ be the transmitted symbols of all the \acp{ut} in the cell over $\tau$ time instances. We can express the received baseband (BB) signal via
	\begin{flalign}
	\label{receivedSigMat}
	\mathbf{Y} = \mathbf{W} \mathbf{H} \mathbf{S}^{T} + \mathbf{W} \mathbf{N},
	\end{flalign}
	where $\mathbf{N} \in \mathbb{C}^{N_{bs} \times \tau}$ represents the \ac{awgn} corrupting the channel output, modeled as having {i.i.d.} zero-mean proper-complex Gaussian entries with variance $p_{n} > 0$. 
	The resulting model is illustrated in Fig. \ref{ModelPicture}.
	
	Our hardware prototype implements the analog combining matrix $\mathbf{W}$. In particular, we consider two different feasible sets for the analog combining weights $\mathcal{W}$:
	\begin{enumerate}[label=\arabic*), leftmargin=2em, labelindent=0pt, itemindent = 0em]
		\item
		\ac{cgac}: such analog combiners can realize any form of analog combining, namely, $\mathcal{W} = \mathbb{C}^{N_{rf} \times N_{bs}}$. This architecture is implemented using a hardware network with controllable gains and phase shifters.
		\item 
		\ac{psoac}: here the elements of the combiner matrix have a fixed unit magnitude. Such analog combiners are implemented using adjustable phase shifters, and thus tend to be less costly and simpler to implement compared to \acp{cgac}. 
	\end{enumerate}
	Since we focus on facilitating channel estimation, we design the analog combining hardware to minimize the \ac{mse} in recovering the channel matrix $\mathbf{H}$ from the observed channel output $\mathbf{Y}$. We assume that $\mathbf{S}$ represents an a-priori known orthogonal pilot sequence, i.e., that channel estimation is carried out in a pilot-aided fashion, where $\mathbf{S} \mathbf{S}^{*} = \mathbf{I}_{\tau}$ and $\tau \ge K$. The design objective can thus be formulated as
	\begin{equation}
	\label{eqn:Objective1}
	\mathbf{W}^{o} = \mathop{\arg \min}\limits_{\mathbf{W} \in \mathcal{W}} \E \left\{\left\|\mathbf{H} - \E\{\mathbf{H} | \mathbf{Y} \} \right\|^2_F \right\}.
	\end{equation}

	We emphasize that the proposed approach can also be extended to multiple cells, in which the channel output is corrupted by an additional interference term, as well as to signal recovery scenarios. In signal recovery, the observed output $	\mathbf{Y} $ in \eqref{receivedSigMat} is used to recover the transmitted symbols $\mathbf{S}$, assuming that knowledge of the channel matrix $\mathbf{H}$ (or a reliable estimate of it) is available.

	\subsection{Channel Model}
	\label{subsec:CM} 
	We consider \ac{mmse} estimation of the unknown channel matrix $\mathbf{H}$ from the channel output $\mathbf{Y}$. We model the distribution of the channel matrix using the common Kronecker model \cite{Costa2010Multiple,bjornson_framework_2010,liu_training_2007}. Accordingly, the matrix $\mathbf{H}$  can be written as   
	\begin{flalign}
	\label{KroneckerModel}
	\mathbf{H} = \mathbf{Q}^{\frac{1}{2}} \bar{\mathbf{H}} \mathbf{P}^{\frac{1}{2}},
	\end{flalign}
	where $\mathbf{Q} \in \mathbb{C}^{N_{bs} \times N_{bs}} $ and $\mathbf{P}\in \mathbb{C}^{K \times K}$ are the deterministic non-singular receive side and transmit side correlation matrices, respectively; and $\bar{\mathbf{H}} \in \mathbb{C}^{N_{bs} \times K}$ models Rayleigh fading, i.e., its entries are {i.i.d.} zero-mean unit variance proper-complex Gaussian \acp{rv}. 
	We henceforth assume that the transmit side correlation matrix is  a scalar multiple of the identity matrix, i.e., $\mathbf{P} = \alpha \mathbf{I}_{K}$ for some $\alpha > 0$, representing the scenario in which the  \acp{ut} are distributed in the cell in an {i.i.d.}. manner.	The correlation matrices $\mathbf{Q} $ and $\mathbf{P} $ are assumed to be known to the \ac{bs}, and can be utilized for recovering the unknown channel $\mathbf{H} $, as discussed in the following section.
	
     Our motivation for focusing on the Kronecker model \eqref{KroneckerModel} with $\mathbf{P} = \alpha \mathbf{I}_{K}$ stems from the fact that this facilitates deriving the optimal unconstrained \ac{cgac}, i.e., the solution to \eqref{eqn:Objective1} when $\mathcal{W} = \mathbb{C}^{N_{rf} \times N_{bs}}$. Nonetheless, our algorithm for designing \acp{psoac}, detailed in Section \ref{sec:Algorithm}, is not restricted to a specific model and only requires an unconstrained \ac{cgac} to approximate. Furthermore, our  hardware prototype detailed in Section \ref{sec:SA} is model independent, and can support any analog combiner configuration.

%----------------------------------------------------------------------------------------
%	RF Chain Reduction Algorithm
%----------------------------------------------------------------------------------------
\section{Analog Combiner Design Algorithm}
\label{sec:Algorithm}
	We now detail the proposed algorithm for designing the analog combining matrix $\mathbf{W}$ based on the objective \eqref{eqn:Objective1}. Our method improves upon the \ac{magiq} algorithm suggested in \cite{Ioushua2019Hybrid}, and consists of two steps: we begin by reformulating the \ac{mse} objective as a matrix trace expression, and then show how this objective can be utilized to design the analog combiner.

	%\vspace{-1em}
	%-----------------------------------
	%	MMSE Channel Estimation
	%-----------------------------------
	\subsection{\ac{mse} Objective}
	\label{subsec:MMSECE}
	In order to specialize the \ac{mse} objective \eqref{eqn:Objective1} to our channel model detailed in Subsection \ref{subsec:CM}, we write the channel input-output relationship \eqref{receivedSigMat} in vector form. In particular, it holds from \eqref{receivedSigMat} that $\mathbf{y} \triangleq {\rm vec}(\mathbf{Y})$ can be written as
	\begin{flalign}
	\label{receivedSigVec}
	\mathbf{y} &= \left(\mathbf{S} \otimes\mathbf{W} \right) \mathbf{h} + \big(\mathbf{I}_{\tau} \otimes\mathbf{W} \big) \mathbf{n},
	\end{flalign}
	where $\mathbf{h} \triangleq {\rm vec}(\mathbf{H}) \sim \mathcal{CN}(\mathbf{0}, \mathbf{P} \otimes \mathbf{Q})$ is the unknown channel in vector form and $\mathbf{n} = {\rm vec}(\mathbf{N}) \sim \mathcal{CN}\big(\mathbf{0}, p_{n} \mathbf{I}_{N_{bs} \cdot \tau}\big)$ is the additive noise. 
	Since $\mathbf{h}$ and $\mathbf{n} $ are mutually independent, it follows from \eqref{receivedSigVec} that $\mathbf{y}$ and  $\mathbf{h}$  are jointly Gaussian. Hence, the \ac{mmse} estimator is given by the linear \ac{mmse} estimator \cite[Ch. 8]{papoulis2002probability}, which can be written as 
	\begin{flalign}
	\nonumber
	\hat{\mathbf{h}} &\triangleq \E\{\mathbf{h} | \mathbf{y}\} = \left(\mathbf{P} \mathbf{S}^{*} \otimes \mathbf{Q} \mathbf{W}^{*} \right) \\
	\label{estimate}
	& \times \left[ \left(\mathbf{S} \mathbf{P} \mathbf{S}^{*} \otimes \mathbf{W} \mathbf{Q} \mathbf{W}^{*} \right) + p_{n} \left(\mathbf{I}_{\tau} \otimes \mathbf{W}\mathbf{W}^{*} \right) \right]^{-1} \mathbf{y}.
	\end{flalign}
	Accordingly, the \ac{mse} $\epsilon \triangleq \E\big\{\|\mathbf{h} - \hat{\mathbf{h}} \|^2 \big\}$ is given by \cite[Ch. 8.4]{papoulis2002probability}
	\begin{flalign}
	\nonumber
	&\epsilon 
	= tr\left(\mathbf{P} \otimes \mathbf{Q} \right) 
	- tr\Big( \left(\mathbf{P} \mathbf{S}^{*} \otimes \mathbf{Q} \mathbf{W}^{*} \right) \big. \Big. \\
	\label{error}
	&\Big. \big. \big[ \left(\mathbf{S} \mathbf{P} \mathbf{S}^{*} \otimes \mathbf{W} \mathbf{Q} \mathbf{W}^{*} \right) 
	+ p_{n} \left(\mathbf{I}_{\tau} \otimes \mathbf{W}\mathbf{W}^{*} \right) \big]^{-1} 
	\left(\mathbf{S} \mathbf{P}^{*} \otimes \mathbf{W} \mathbf{Q}^{*} \right) \Big).
	\end{flalign}
%	\begin{flalign}
%	\nonumber
%	\epsilon =& tr\left(\mathbf{P} \otimes \mathbf{Q} \right) 
%	- tr\Big( \left(\mathbf{P} \mathbf{S}^{*} \otimes \mathbf{Q} \mathbf{W}^{*} \right) \big. \Big. \\
%	\nonumber
%	&\qquad \qquad \quad \Big. \big. \big[ \left(\mathbf{S} \mathbf{P} \mathbf{S}^{*} \otimes \mathbf{W} \mathbf{Q} \mathbf{W}^{*} \right) 
%	+ p_{n} \left(\mathbf{I}_{\tau} \otimes \mathbf{W}\mathbf{W}^{*} \right) \big]^{-1} \big. \Big. \\
%	\label{error}
%	&\qquad \qquad \qquad \qquad \Big. \big. \left(\mathbf{S} \mathbf{P}^{*} \otimes \mathbf{W} \mathbf{Q}^{*} \right) \Big).
%	\end{flalign}
	The \ac{mse} \eqref{error} is determined by the pilot symbols $\mathbf{S}$, the second order statistical moments of the channel, represented by the correlation matrices $\mathbf{Q}, \mathbf{P}$, and the analog combiner $\mathbf{W}$. We next assume that $\mathbf{S}$, $\mathbf{P}$, and $\mathbf{Q}$ are known and seek the optimal combiner $\mathbf{W}$ that minimizes the \ac{mse} of \eqref{error}.
	%For a-priori known and fixed $\mathbf{S}$, $\mathbf{P}$, and $\mathbf{Q}$,  the design of the analog combiner under the \ac{mse} objective \eqref{error} is detailed in the following subsection. 

	%-----------------------------------
	% Analog Combiner Design
	%-----------------------------------
	\subsection{Analog Combiner Design}
	\label{subsec:ACD}

	Minimization of \eqref{error} is equivalent to maximization of the second trace term of \eqref{error}, i.e.,
	\begin{flalign}
	\label{maximization}
	\mathbf{W}^{o} = \mathop{\arg\max}\limits_{{\mathbf{W}\in \mathcal{W}}} \quad f( \mathbf{W} ) 
	\end{flalign}
	where
	\begin{flalign}
	\nonumber
	f( \mathbf{W} ) 
	&= tr\Big( \left(\mathbf{P} \mathbf{S}^{*} \otimes \mathbf{Q} \mathbf{W}^{*} \right) \big[ \left(\mathbf{S} \mathbf{P} \mathbf{S}^{*} \otimes \mathbf{W} \mathbf{Q} \mathbf{W}^{*} \right) \big. \Big. \\
	\label{FW}
	&+ \Big. \big. p_{n} \left(\mathbf{I}_{\tau} \otimes \mathbf{W}\mathbf{W}^{*} \right) \big]^{-1} \left(\mathbf{S} \mathbf{P}^{*} \otimes \mathbf{W} \mathbf{Q}^{*} \right) \Big).
	\end{flalign}
	Let $\mathcal{U}$ and $\mathcal{D}$ be the sets of unitary ${N_{rf} \times N_{rf}}$ matrices and diagonal ${N_{rf} \times N_{rf}}$ matrices with positive diagonal entries. The analog combiner matrix $\mathbf{W}^o$ for unconstrained analog combiners, i.e., $\mySet{W} = \mathbb{C}^{N_{rf} \times N_{bs}}$, is given in the following theorem:
	\begin{theorem}
		\label{thm:unconst}
		Let $\mathbf{U}^o$ be an $N_{bs} \times N_{rf}$ matrix whose columns are the $N_{rf}$ eigenvectors of $\bar{\mathbf{Q}}$ corresponding to its  $N_{rf}$ largest eigenvalues, where 
		\begin{flalign}
		\label{optimalU}
		\bar{\mathbf{Q}} 
		\triangleq\left( \alpha \mathbf{Q} + p_{n}\mathbf{I}_{N_{bs}} \right)^{-\frac{1}{2}} (\alpha \mathbf{Q})^{2} \left( \alpha \mathbf{Q} + p_{n}\mathbf{I}_{N_{bs}} \right)^{-\frac{1}{2}}.
		\end{flalign}
		Then, for any  $\mathbf{V}  \in \mathcal{U}$ and  $\mathbf{D}  \in \mathcal{D}$, the optimization problem \eqref{maximization} is solved by setting 
		\begin{flalign}
		\label{optimalW}
		\mathbf{W}^{o} =  \mathbf{V} \mathbf{D} \big(\mathbf{U}^o\big)^{*}.
		\end{flalign}
	\end{theorem}
	
	\begin{IEEEproof}
		The proof is detailed in Appendix \ref{theorem1}.
	\end{IEEEproof}
	We note that in the high SNR regime, i.e., when $p_n \approx 0$, $\mathbf{U}$ in Theorem \ref{thm:unconst} becomes the first $N_{rf}$ eigenvectors of $\alpha \mathbf{Q}$, coinciding with the derivation in \cite[Sec. IV]{Ioushua2018Pilot} which assumed a noiseless setup.

	The analog combiner in \eqref{optimalW} is achievable for any  $\mathbf{V}$ and $\mathbf{D}$ using the \ac{cgac} architecture, in which each element of the analog combining matrix can be any complex value. For the \ac{psoac} case, the entries of the combiner matrix are restricted to have unit amplitude, a condition which may not be satisfied for a matrix of the form \eqref{optimalW}. Following \cite{Ioushua2019Hybrid}, we propose to exploit the non-uniqueness  of $\mathbf{W}^{o}$ to facilitate its approximation using a feasible \ac{psoac} matrix. In particular, we recover the selection of the non-unique $\mathbf{V}$ and $\mathbf{D}$ for which the resulting   $\mathbf{W}^{o}$ can be closely approximated using a feasible $\mathbf{W}$, namely, our design objective is
	\begin{flalign}
	\label{problem}
	\mathbf{W}^{po}= \mathop{\arg\min}\limits_{\mathbf{W} \in \mySet{W}, \mathbf{V} \in \mathcal{U}, \mathbf{D} \in \mathcal{D}} &\quad \| \mathbf{W} - \mathbf{V} \mathbf{D} \big(\mathbf{U}^o\big)^{*} \|_F^{2}.
	\end{flalign}
	In \cite[Sec. V]{Ioushua2019Hybrid}, only the non-uniqueness in the unitary $\mathbf{V}$ is exploited, and the diagonal $\mathbf{D}$ is assumed to be the identity matrix $\mathbf{I}_{N_{rf}}$. Consequently, our proposed design criterion generalizes that of \ac{magiq} \cite{Ioushua2019Hybrid}, and is capable of recovering \ac{psoac} matrices which better approximate the unconstrained \ac{mse} minimizing analog combiner compared to \ac{magiq} \cite{Ioushua2019Hybrid}. 
	
	We tackle the optimization problem \eqref{problem} in an alternating fashion. Our design method is based on the following lemma:
	\begin{lemma}
		\label{lem:Alt}
		For any fixed $\mathbf{V} \in \mathcal{U}$ and $\mathbf{D} \in \mathcal{D}$, if $\mathcal{W}$ represents  \ac{psoac} matrices then
		\begin{subequations}
			\label{eqn:Alt}
			\begin{align}
			\tilde{\mathbf{W}} 
			&= \mathop{\arg\min}\limits_{\mathbf{W} \in \mySet{W}} \| \mathbf{W} - \mathbf{V} \mathbf{D} \mathbf{U}^{*} \|_F^{2} 
			= \mathcal{P}( \mathbf{V} \mathbf{D} \mathbf{U}^{*}).
			\label{solutionW}
			\end{align}
			where $[\mathcal{P}(\mathbf{U})]_{ij} \triangleq e^{j 2 \pi \angle [\mathbf{U}]_{ij}}$.
			Furthermore, for any $\mathbf{W} \in \mathcal{W}$ and $\mathbf{D} \in \mathcal{D}$, by letting $\mathbf{L}$ and $\mathbf{R}$ be the left and right singular vectors matrices of $\mathbf{W}  \mathbf{U} \mathbf{D}$, respectively, it holds that
			\begin{align}
			\tilde{\mathbf{V}} 
			&= \mathop{\arg\min}\limits_{\mathbf{V} \in \mySet{U}} \| \mathbf{W} - \mathbf{V} \mathbf{D} \mathbf{U}^{*} \|_F^{2} 
			= \mathbf{L}\mathbf{R}^*.
			\label{solutionV}
			\end{align}
			Finally,  letting $\eta > 0$ be some lower bound on the diagonal entries of the matrices in $\mathcal{D}$, guaranteeing that these values are strictly positive, it holds that for any $\mathbf{W} \in \mathcal{W}$ and $\mathbf{V} \in \mathcal{U}$, the diagonal entries of  $  \tilde{\mathbf{D}} = \mathop{\arg\min}\limits_{\mathbf{D} \in \mySet{D}} \| \mathbf{W} - \mathbf{V} \mathbf{D} \mathbf{U}^{*} \|_F^{2}  $ are given by
			\begin{equation}
			\label{solutionD}
			\big(\tilde{\mathbf{D}}\big)_{l,l} = 
			\max \left( \frac{ {\rm Re} \left( \left( \mathbf{W}^{*} \mathbf{V} \right)_{:,l}^{*} \left( \mathbf{U} \right)_{:,l} \right) }{ \| \left( \mathbf{U} \right)_{:,l} \|^{2} }, \eta \right), 
			\end{equation}
			for all  $j = 1,\ldots, N_{bs}$.
		\end{subequations}
	\end{lemma}
	
	\begin{IEEEproof}
		The lemma directly follows from \cite[Lem 2]{shlezinger2019dynamic}.
	\end{IEEEproof}
	
	\smallskip
	Lemma \ref{lem:Alt} implies that the optimization problem \eqref{problem} can be solved using alternating optimization. In particular, we propose to update each of the three matrices $\mathbf{W}$, $\mathbf{V}$, and $\mathbf{D}$ in turn, while fixing the remaining two matrices, and to repeat this process iteratively. Since our objective in \eqref{problem} is the minimization of the convex Frobenius norm, it follows from \cite[Thm. 2]{bezdek2003convergence} that the convergence of such an alternating approach is guaranteed. The proposed iterative alternating optimization method is summarized in Algorithm \ref{alg:algorithm}.
    For a given $\mathbf{U}^o$, MaGiQ \cite{Ioushua2019Hybrid} can be considered as a special case of the proposed algorithm with $\mathbf{D}$ fixed to the identity matrix, i.e., without step \ref{stp:Dmat}. This additional degree of freedom allows our algorithm to obtain close feasible approximations of the unconstrained optimal analog combiner $\mathbf{W}^o$, at the cost of the additional computation of the $N_{rf}$ diagonal entries of  $\mathbf{D}$ in each iteration via \eqref{solutionD}.

		\begin{algorithm}[!th]
		\caption{Iterative alternating algorithm for \ac{psoac} matrices}
		\label{alg:algorithm}
		\begin{algorithmic}[1]
			%\Procedure{IAA}{$input1,input2$}
			\State \textbf{Input:} 
			Receive side correlation matrix $\mathbf{Q}$.
			%$\mathbf{T}_{i}$ and $\mathbf{D}_{i}$.
			\State \textbf{Output:} 
			$\mathbf{W}^{po} \in \mathcal{W}$ (with unit magnitude entries).
			\vspace{0.2em}
			\State \textbf{Initialization:} 
			$i := 0$, $\mathbf{V}_{i} := \mathbf{I}_{N_{rf}}$ and $\mathbf{D}_{i} := \mathbf{I}_{N_{rf}}$.
			\State Compute $\mathbf{U}$ as the first $N_{rf}$ eigenvectors of $\bar{\mathbf{Q}}$ in \eqref{optimalU}.
			\While{termination criterion is inactive}
			\State Obtain $\mathbf{W}_{i+1}$ via \eqref{solutionW} with $\mathbf{V} = \mathbf{V}_i$ and $\mathbf{D} = \mathbf{D}_i$.
			\State Obtain $\mathbf{V}_{i+1}$ via \eqref{solutionV}  with $\mathbf{W} = \mathbf{W}_{i+1}$ and $\mathbf{D} = \mathbf{D}_i$.
			\State \label{stp:Dmat} Obtain $\mathbf{D}_{i+1}$ via \eqref{solutionD} with $\mathbf{W} = \mathbf{W}_{i+1}$ and $\mathbf{V} = \mathbf{V}_{i
				+1}$.
			\State $i := i + 1$.
			\EndWhile\label{algoendwhile}
			\State \textbf{return} $\mathbf{W}^{po} = \mathbf{W}_i$.
		\end{algorithmic}
	    \end{algorithm}
	
\begin{figure}[t!]
		\centering
		\vspace{-1em}
		\includegraphics[height=5.8cm, width=8.6cm]{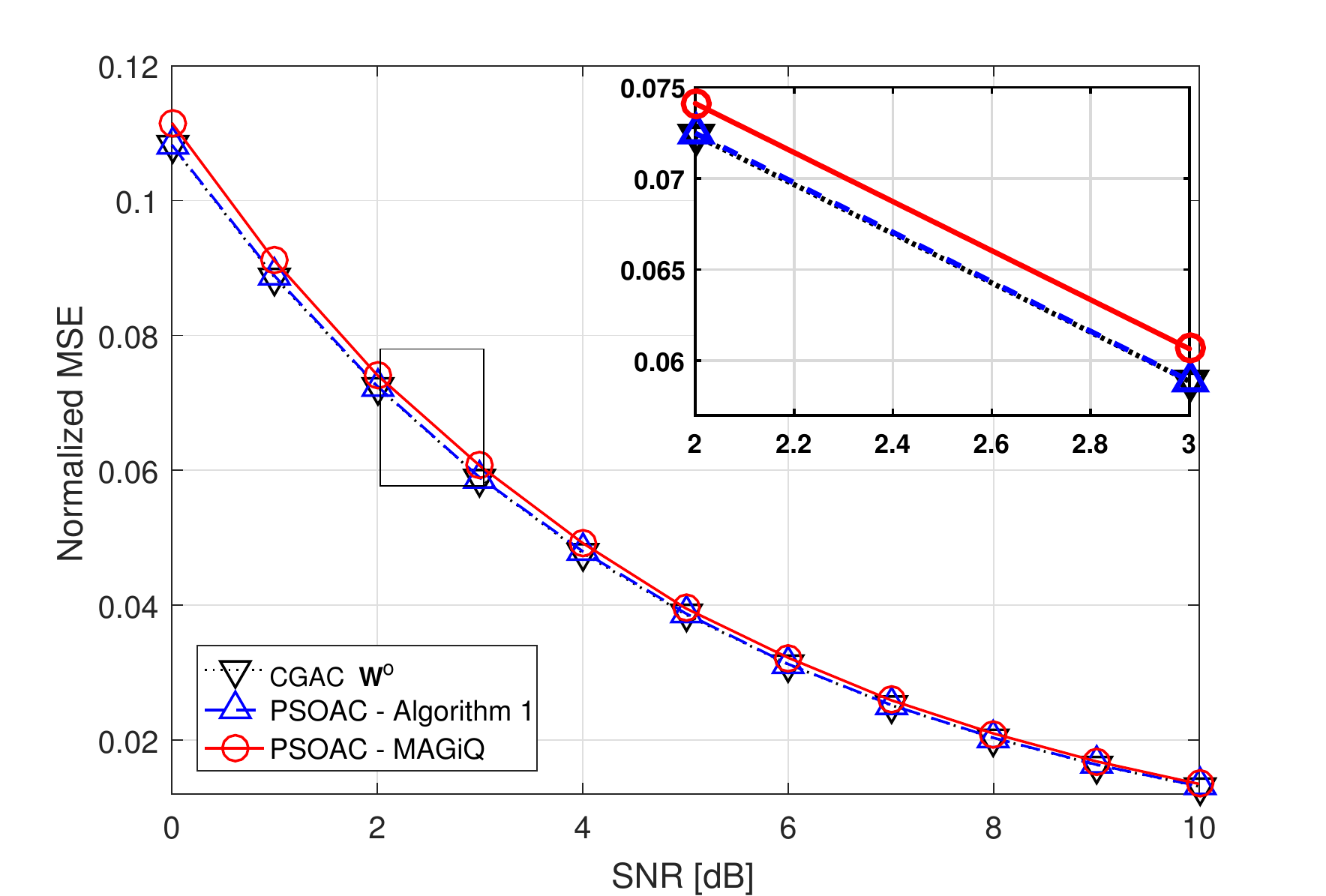}
		\vspace{-0.6em}
		\caption{Normalized \ac{mse} versus \ac{snr}, $N_{bs} = 80$, $N_{rf} = 20$.}
		\label{fig:Network1}
		\vspace{0.8em}
	\end{figure}

	The fact that our algorithm obtains a better approximation of $\mathbf{W}^o$ compared to \ac{magiq} also translates to improved channel estimation accuracy, as demonstrated in the following example. Consider a multi-user \ac{mimo} system in which a \ac{bs} equipped with $N_{bs} = 80$ antennas and $N_{rf} = 20$ RF chains serves $K=40$ \acp{ut}. The receive side correlation matrix $\mathbf{Q}$ follows the Jakes' model with antenna spacing of $0.2$ carrier wavelength \cite{jakes1994microwave}, and the transmit side correlation $\mathbf{P}$ is set to $\mathbf{I}_{K}$. 
	In Fig. \ref{fig:Network1} we evaluate the  normalized \ac{mse}, defined as $\frac{\epsilon}{ \E\big\{\|\mathbf{h} \|^2 \big\}}$, versus \ac{snr}, given by ${\rm SNR} = p^{-1}_n$. The normalized \ac{mse}s are computed using the unconstrained \ac{cgac} $\mathbf{W}^o$ as well as \acp{psoac} designed using \ac{magiq} and the proposed algorithm, respectively.  %The termination criterion used in Algorithm \ref{alg:algorithm} is called when the iterative process either converges or when the iterations are exhausted. 
	Observing Fig. \ref{fig:Network1} we note that the \ac{psoac} designed using the proposed algorithm achieves effectively the same performance as the \ac{cgac} $\mathbf{W}^o$ which requires controllable gains, while the normalized \ac{mse} achieved using \ac{magiq} is within a small gap from $\mathbf{W}^o$. This small gap in normalized \ac{mse} can lead to substantial gaps in \ac{mse}, particularly when the Frobenius norm of the channel,  $\E\big\{\|\mathbf{h} \|^2 \big\}$, is large, as common in massive \ac{mimo} systems. This numerical study demonstrates the superiority of the proposed algorithm over previous \ac{psoac} design methods. Consequently, our hardware prototype and its experimental system use Algorithm \ref{alg:algorithm} when realizing \acp{psoac}.
	
	In our prototype we implement both \acp{cgac}, designed via \eqref{optimalW} by setting $\mathbf{V} = \mathbf{D} = \mathbf{I}_{N_{rf}}$, and \acp{psoac}, configured using Algorithm \ref{alg:algorithm}.  The architecture of the prototype, which allows it to implement the aforementioned hybrid design in a dynamic manner, is discussed in the following section.

	\begin{figure*}[t!]
		\centering
		\includegraphics[height=6.8cm, width=16.6cm]{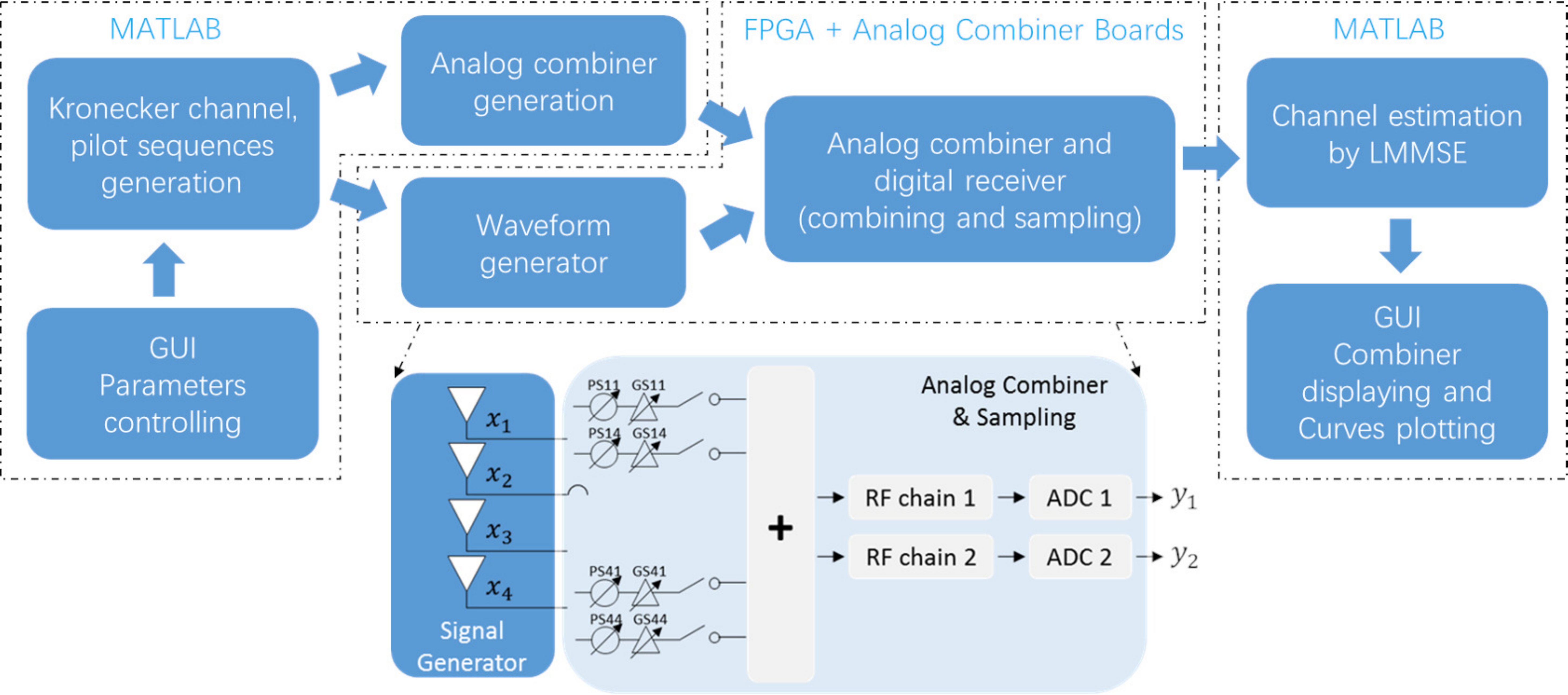} \\
		\vspace{0.2cm}
		\includegraphics[height=4.1cm, width=16.6cm]{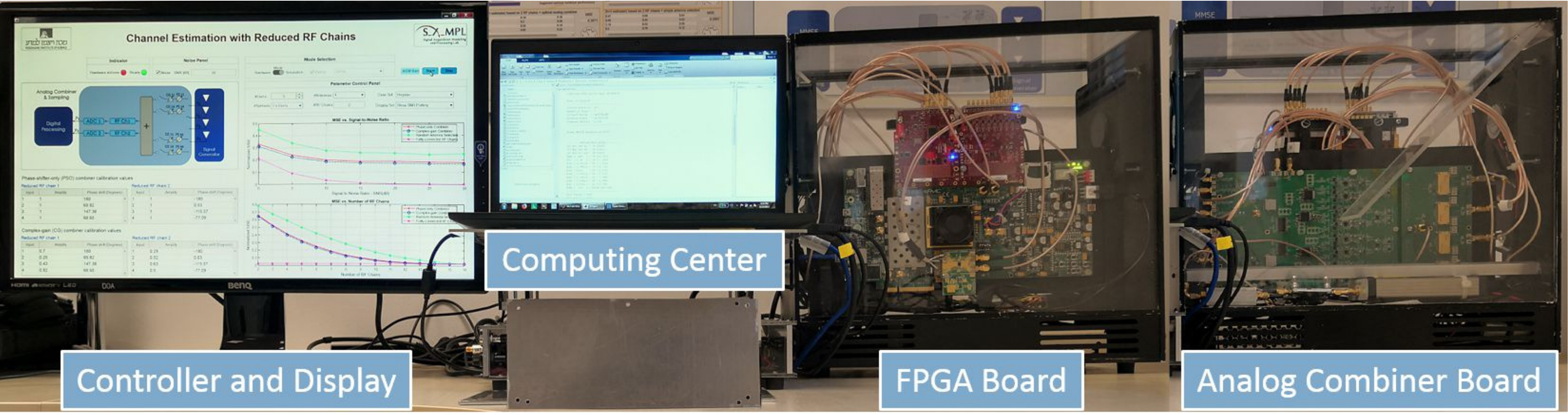}
		\caption{Top: overview of the information flow in the experimental setup. Bottom: overview of the prototype and its subsystems.}
		\label{Prototype}
	\end{figure*}

%----------------------------------------------------------------------------------------
%	System Architecture
%----------------------------------------------------------------------------------------
\smallskip
\section{System Architecture}
\label{sec:SA}
	In this section, we elaborate on the system architecture of the hardware prototype which realizes the RF chain reduction scheme detailed in the previous section. To that aim, we first present the high-level system architecture in Subsection \ref{subsec:DP}, after which we discuss the concrete structure of each of the hardware components in Subsection \ref{subsec:PPE}.

	%----------------------------------------------------------------------------------------
	%	Design Philosophy
	%----------------------------------------------------------------------------------------
	%\vspace{-0.2cm}
	\subsection{High-Level Design}
	\label{subsec:DP}

	\subsubsection{Experimental Environment}
	\label{subsubsec:EE}
	Our hardware prototype implements a configurable analog combiner, which is evaluated using a dedicated experimental setup at microwave frequencies.  
	This setup consists of a Matlab-based host application and a \ac{fpga} board. The former simulates the BB channel output and processes the signal captured after analog combining. The latter acts as an interface between the digital signals generated and processed by the host application, and the analog signals which are utilized by the analog combiner hardware.
	In particular, the input and output signals of the analog combiner hardware are generated as follows:
	\begin{itemize}[leftmargin=2em, labelindent=0pt, itemindent = 0em]
		\item  
		Analog combiner input: The digital baseband channel outputs simulated by the host application are transferred by an Ethernet cable from the host application to the \ac{fpga} board in real-time. The \ac{fpga} board generates the baseband input signal which is up-converted on the combiner board using a carrier waveform generated by a VSG25A vector signal generator. The resulting analog passband signal represents the multivariate channel output observed at the \ac{bs} antenna array. 
		\item  
		Analog combiner output: The analog combined passband signal, representing the signal fed to the RF chains at the \ac{bs}, is down-converted with the same carrier waveform as for the up-conversion on the combiner board, followed by an analog-to-digital conversion implemented on the \ac{fpga}.  These digital outputs are transferred from the \ac{fpga} board to the host application where they are utilized for estimating the underlying channel. 
	\end{itemize}
	The host application is also capable of adjusting the weights of the analog combining hardware via the \ac{fpga} board. 
	The experimental procedure is illustrated in the flow chart at the top of Fig. \ref{Prototype}.

\subsubsection{Analog combiner implementation}
\label{subsubsec:ACI} 
	Several different architectures for analog combiners can be found in the literature \cite{Méndez2016Hybrid, Ioushua2019Hybrid}: The most common is arguably a controllable network of fully connected phase shifters, i.e., the \ac{psoac}. Alternative architectures include fully connected complex gain networks, such as the \ac{cgac};  fully connected phase shifters and switches network, and flexible partially connected phase shifters network with sub arrays. We refer the readers to \cite[Sec. II]{Méndez2016Hybrid} for a detailed account of these analog combining architectures. In order to incorporate a large family of architectures, our analog combiner hardware consists of a controllable network of gains and phase shifters.  The outputs of the adjustable gains and phase shifters are then summed by a combiner and fed to an RF chain. In particular, our hardware system consists of 4 input ports and 2 output ports, namely, it can be utilized in  a \ac{bs} with 4 antennas and 2 RF chains. This setup can also be used for experimenting with analog combiners with larger number of antennas and RF chains using a virtual channel extension. 
	
	By using the flexible and controllable network of gains and phase shifters, we implement the proposed designs for both CGACs as well as PSOACs. The baseline is to use the hardware board which models a \acs{mimo} \acs{bs} with 4 antenna inputs and 2 RF chains. While this combiner implements a hybrid MIMO receiver with $N_{bs} =4$ and $N_{rf} = 2$, it can also be used, along with the experimental setup detailed in the sequel, to evaluate in hardware hybrid MIMO BS with 8 antenna inputs and 4 RF chains as well as 16 antenna inputs and 8 RF chains. The ability to use the board and the experimental setup to evaluate hybrid receivers with larger number of antennas and RF chains is obtained using a virtual channel extension. This virtual channel approach is based on a sequential utilization of the basic hardware to obtain an overall combined result.

	To present the virtual channel extension, we note that  the two outputs of the analog combiner are obtained as a linear combination of four antenna inputs with different weights, which are determined by some basic combining matrix $\mathbf{W}_b$. Defining $\mathbf{X}_b \triangleq \mathbf{H}\mathbf{S}^{T} + \mathbf{N}$, we denote the channel output with this basic combining as
	\begin{flalign}
	\label{BasicLC}
	\mathbf{Y}_{b} = \mathbf{W}_{b} \mathbf{X}_{b}.
	\end{flalign}
	In order to simulate more antenna inputs and RF chains, we utilize multiple time instances to form a single analog combining input-output pair. 
	For example, in order to realize  8 antenna inputs and 4 RF chain, we use two different $4 \times 1$ channel output vectors, denoted $\mathbf{X}_{b}^1$ and $\mathbf{X}_{b}^2$, and four $4 \times 2$ analog combining matrices $\mathbf{W}_{b}^{11}, \mathbf{W}_{b}^{21}, \mathbf{W}_{b}^{12}, \mathbf{W}_{b}^{22} $. Now, following  \eqref{BasicLC} we can obtain a configurable $8 \times 4$ analog combiner via the assignment
	\begin{flalign}
	\label{AdvancedLC}
	\left[\begin{array}{c}
	\mathbf{Y}_{b}^{1} \\
	\hline
	\mathbf{Y}_{b}^{2}
	\end{array} \right]
	=\left[\begin{array}{c|c}
	\mathbf{W}_{b}^{11} & \mathbf{W}_{b}^{12} \\
	\hline
	\mathbf{W}_{b}^{21} & \mathbf{W}_{b}^{22}
	\end{array} \right]
	\left[ \begin{array}{c}
	\mathbf{X}_{b}^{1} \\
	\hline
	\mathbf{X}_{b}^{2}
	\end{array} \right].
	\end{flalign}
	It follows from  \eqref{AdvancedLC} that we need to sequentially utilize the basic hardware four times to complete the divided four blocks of the analog combiner matrix. Following the same approach as the realization of 8 antenna inputs and 4 RF chains,   a setup with 16 antenna inputs and 8 RF chains can be obtained  by utilizing 16 different $4 \times 2$ basic combiners  to complete the divided 16 blocks of the analog combiner matrix. Clearly, analog combiners with different  analog combining ratios, namely, in which the number of antennas is not twice the number of RF chains, can be realized following the same guidelines by using only the required number of input and output ports.

	%----------------------------------------------------------------------------------------
	%	Prototype Physical Entity
	%----------------------------------------------------------------------------------------
	\begin{table*}[!t]
		\footnotesize
		\renewcommand{\arraystretch}{1.2}
		\caption{Controllable parameters supported by \ac{gui}}
		\label{table1}
		\centering
		\begin{tabular}{c|c|c|c}
			\hline
			Working mode & \multicolumn{2}{|c|}{Simulation} & Hardware \\
			\hline
			Curve display mode & NMSE vs. SNR & NMSE vs. $N_{rf}$ & NMSE vs. SNR \\
			\hline 
			Number of UTs & \multicolumn{3}{|c}{$K = 1, \ldots, 10$} \\
			\hline 
			Number of training symbols & \multicolumn{3}{|c}{$\tau = 1 \times K$, $\tau = 2 \times K$, $\tau = 3 \times K$} \\
			\hline 
			Number of receiving antennas in the BS & \multicolumn{3}{|c}{$N_{bs} = 4$, $N_{bs} = 8$, $N_{bs} = 16$} \\
			\hline 
			Number of RF chains in the BS & $\frac{N_{bs}}{2}$ & $2, \ldots, N_{bs}$ & $\frac{N_{bs}}{2}$ \\
			\hline 
			Rank of receive side correlation matrix & \multicolumn{3}{|c}{Regular case: $N_{rf} < rank \le N_{bs}$, Best case: $rank = N_{rf}$} \\
			\hline 
		\end{tabular}
	\end{table*}

	\begin{figure*}[t!]
		\centering
		\includegraphics[height=10.2cm, width=16.0cm]{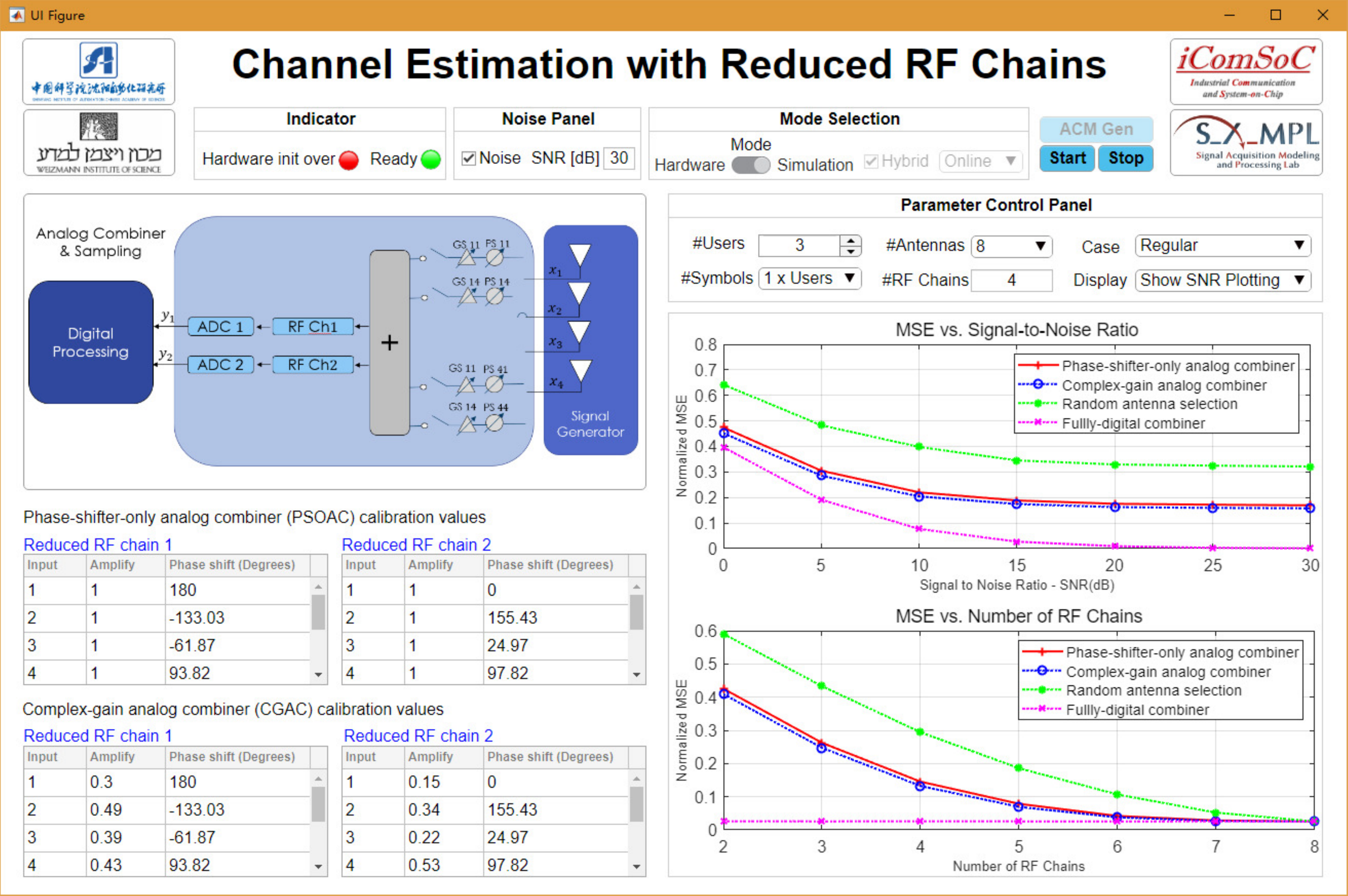}
		\caption{Overview of the GUI}
		\label{GUI}
	\end{figure*}

	\begin{figure*}[t!]
		\centering
		\includegraphics[height=11.6cm, width=16.6cm]{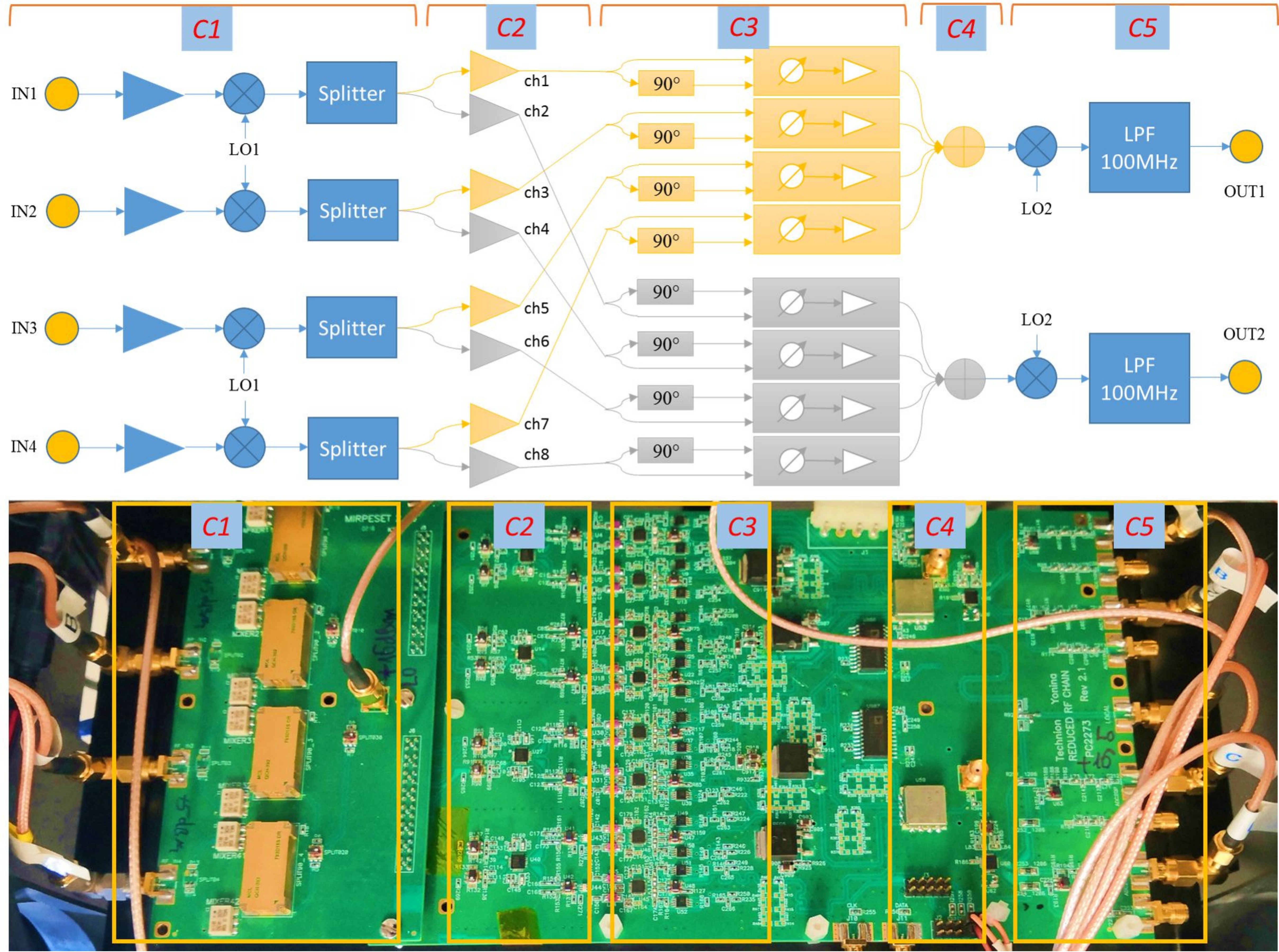}
		\caption{Overview of the analog combiner board. Top: block diagram. Bottom: circuit board.}
		\label{AnalogCombiner}
	\end{figure*}

\subsection{Prototype Physical Entities}
\label{subsec:PPE}  
	The overall prototype system is depicted at the bottom of Fig. \ref{Prototype}. The prototype consists of the following components: a controller and display, providing the \ac{gui}, a computing center running the Matlab-based host application, an \ac{fpga} board, and the analog combiner hardware. In the following we elaborate on each of these blocks.

\subsubsection{Controller and display}
\label{subsubsec:CD}
	The \ac{gui} allows to configure and evaluate the experimental setup in a user-friendly environment. In particular, our \ac{gui} provides the ability to change the main parameters of the experiment and to numerically compare the normalized \acp{mse} obtained in two display modes: with respect to \ac{snr} or number of RF chains.
	The main controllable parameters include the number of \acp{ut}, training symbols, receive antennas, and the rank of the receive side correlation matrix. Details of the supported parameter combinations are summarized in Table \ref{table1}. 
	
	Once the experimental setup is configured and a test is launched, the \ac{gui} presents in real-time the selected values used to configure the analog combining hardware for both \ac{psoac} as well as \ac{cgac}. 
	Furthermore, the \ac{gui} presents updated normalized \ac{mse} curves during the test and after it is concluded, comparing the performance of the utilized \ac{psoac} and \ac{cgac} to analog combiners in which the RF chains are directly connected to randomly selected antennas, as well as to a fully-digital setup. The fully-digital setup, representing the performance achievable without analog combining when each antenna feeds a dedicated RF chain, constitutes a fundamental lower bound on the channel estimation \ac{mse} with RF chain reduction. An overview of the GUI is depicted in Fig. \ref{GUI}.

\subsubsection{Computing Center}
\label{subsubsec:CC}
	The computing center is a 64-bit laptop with 4 CPU cores and 16GB RAM running the Matlab-based host application. The application is controlled by the \ac{gui}, and implements the following functionalities: 
	\begin{itemize}[leftmargin=2em, labelindent=0pt, itemindent = 0em]
		\item
		The host application computes the analog combiner weights using the algorithm detailed in Section \ref{sec:Algorithm}, and adjusts the analog combiner hardware before each simulation test.
		\item
		The application generates the digital baseband signals, i.e., the  pilot sequences, as well as the wireless channel outputs, which are fed to the \ac{fpga} to generate the analog combiner input.
		\item
		On the receive side, the application processes the baseband channel outputs and produces the \ac{mmse} channel estimate via \eqref{estimate}.  
	\end{itemize}
	The communication  between the computing center and the hardware board is carried out over an Ethernet cable. Through the cable, the generated digital baseband channel outputs and analog combiner matrix are transmitted to the build-in memory of the \ac{fpga} board, and the baseband RF chain outputs are acquired.

\subsubsection{FPGA board}
\label{subsubsec:FB}
	The FPGA board consists of an off-the-shelf Xilinx VC707 evaluation board. The evaluation board utilizes a 4DSP FMC204 16-bit DAC mezzanine card for baseband waveform generation, as well as an eight-channel 4DSP FMC168 16-bit digitizer  card for sampling of the combined analog signal.
	
	In the baseband analog signal generation process, the waveforms are stored as digital baseband I/Q pairs on the build-in memory. Then, the FPGA device reads out the pre-stored waveforms from the memory and employs an $8$ Gbps Serializer/Deserializer (SerDes) device to transfer it to the $16$-bit DAC mezzanine card. The DAC card then interpolates and converts the stored waveforms to analog baseband signals at a sample rate of $250$ Msps. The analog baseband signals are transmitted to the analog combiner board through coaxial cables where they are up-converted to passband and linearly combined.
	
	The analog combiner outputs, which are down-converted to baseband on the combiner board, are digitized using four out of eight channels of the $16$-bit digitizer card with a $250$ MHz sampling rate. Each I/Q pair occupies two channels of the digitizer card. The $1.2$ Gbps SerDes transfers the sampled data to the FPGA who then writes the data to a digital first-in-first-out (FIFO) buffer for reading by the computing center.
	
	The FPGA also produces selection and control commands for configuring of analog combiner. Once the host application produces an analog combiner configuration, the weights matrix is  provided  to the FPGA board via the FIFO buffer. These weights are then transferred to the analog combiner hardware using a serial peripheral interface (SPI) protocol, which is used to control the dedicated analog combiner board.

\subsubsection{Analog combiner board}
\label{subsubsec:ACB}
	The analog combiner board is a self-designed dedicated hardware, which realizes a controllable analog combiner network, and can serve as any of the common combiner architectures, e.g., phase shifter networks, switching networks \cite{Ioushua2019Hybrid} and DFT beamforming \cite{kim2015mse}.  The designed analog combiner supports two different types of input signals: one is passband signals in the frequency band up to $4.5$ GHz and another is BB signals each with a $125$ MHz maximum bandwidth. In our experimental setup we use BB signals as our inputs, where up-conversion and down-conversion are carried out on the analog combiner board.
	The board consists of five different blocks, implementing the main different functionalities, namely: 
	\begin{enumerate}[label={\em C\arabic*}, leftmargin=3em, labelindent=0pt, itemindent = 0em]
		\item \label{itm:1} 
		Up-conversion of incoming signal and signal splitting.
		\item \label{itm:2} 
		Passband signal input and amplification.
		\item \label{itm:3} 
		Weights (phases and gains) application and configuration.
		\item \label{itm:4} 
		Summing up of the incoming signals.
		\item \label{itm:5} 
		Down-conversion and low-pass filtering of combined signals.
	\end{enumerate}
	The individual blocks of the analog combiner board are marked in Fig. \ref{AnalogCombiner}, which depicts the block diagram of the hardware as well as the circuit board. 
	
	In block \ref{itm:1}, each of the four inputs is a complex BB signal transmitted from the \ac{fpga}, feeding an amplifier followed by a mixer. The BB signals whose maximal range of frequency is $125$ MHz are up-converted to RF signals at a frequency of $1$ GHz. The RF signals represent the passband signals observed by the \ac{bs} antennas and each power level of the RF signals are tuned to $0$ dBm and split into two same signals. In order to support passband inputs, one must only replace block \ref{itm:1} with the corresponding passband signals observed by the antennas.
	
	Each of the passband signals is forwarded to block \ref{itm:2}, where it is fed to an amplifier with $16$ dB fixed gain.
    Specifically, we use an ADL5566 dual RF operational amplifier  for each splitted pair of passband signals. This amplifier has a low latency and a very flat frequency response,
    including a channel-to-channel gain and phase errors of $0.1$ dB and $0.06$ degree, respectively, at $100$ MHz. Furthermore, the ADL5566 supports input signals with up to $4.5$ GHz bandwidth without having the amplification vary over frequency, while hardly affecting the \ac{snr} due to its low noise input stage of only $1.3nV/\sqrt{Hz}$. 
	Thus, the receiver contains four ADL5566 amplifiers for the four RF input signals. The wideband range property of ADL5566 makes our hardware suitable for a broad range of RF signals used in actual communication systems.
	
	The signals from the amplifiers are separated into two groups. Each group has four inputs, including one signal from each splitted pair. Then, each signal is splitter again into  two signals with $90$-degree offset, which are used as the inputs to a  phase shifter and gain block. The purpose of this additional splitting with phase offset stems from the fact that we use differential phase shifters which operate on such inputs. In block \ref{itm:3}, each of the four channels in one of the groups is activated by an ADL5390 analog vector multiplier, implementing the phase and gain of each analog combining weight applied to the input signal. 
	The weights applied in the vector multiplications are determined by the output DC level of an AD7808 octal 10-bit DACs with serial load capabilities, which receives control commands from the FPGA board used to configure the analog combining weights by finding a look-up table. 
	The usage of controllable gains and phases requires a calibration stage when the interconnections are established to guarantee that the configured weights are correctly translated  into the desired phase and gain values.
	
	Arguably, the most common analog combiner architecture is based on phase shifters \cite{Méndez2016Hybrid,mo2017hybrid,roth2018comparison}. In practice, applied phase shifters are digitally controlled with phase resolutions typically above $5$ degrees. This crude resolution may significantly degrade the system performance by inducing quantization errors. In our novel combiner prototype, we used vector multiplexer ADL5390 as an analog phase shifter controlled by the 10-bit DAC, allowing to realize combiner with controllable gains as well as providing the ability to reach an improved resolution of less than $1.5$ degree in phase.
	
	The outputs of each group are summed in block \ref{itm:4}, which finalizes the realization process of analog combining and obtains a combined passband signal. Then, the combined signals are down-converted in block \ref{itm:5}  using a set of ADL5382 down-converters with the same local oscillator used for up-conversion. The down-converted signals are filtered to baseband signals with a maximum $125$ MHz bandwidth.
	Finally, the signals of the two outputs are forwarded to the \ac{fpga} where they are acquired using a 4DSP FMC168 $16$-bit digitizer card, obtaining the digital outputs for further digital signal processing in the computing center.

%----------------------------------------------------------------------------------------
%	Experiment Results
%----------------------------------------------------------------------------------------
\begin{figure*}[t!]
	\centering
	\includegraphics[height=9.8cm, width=16.8cm]{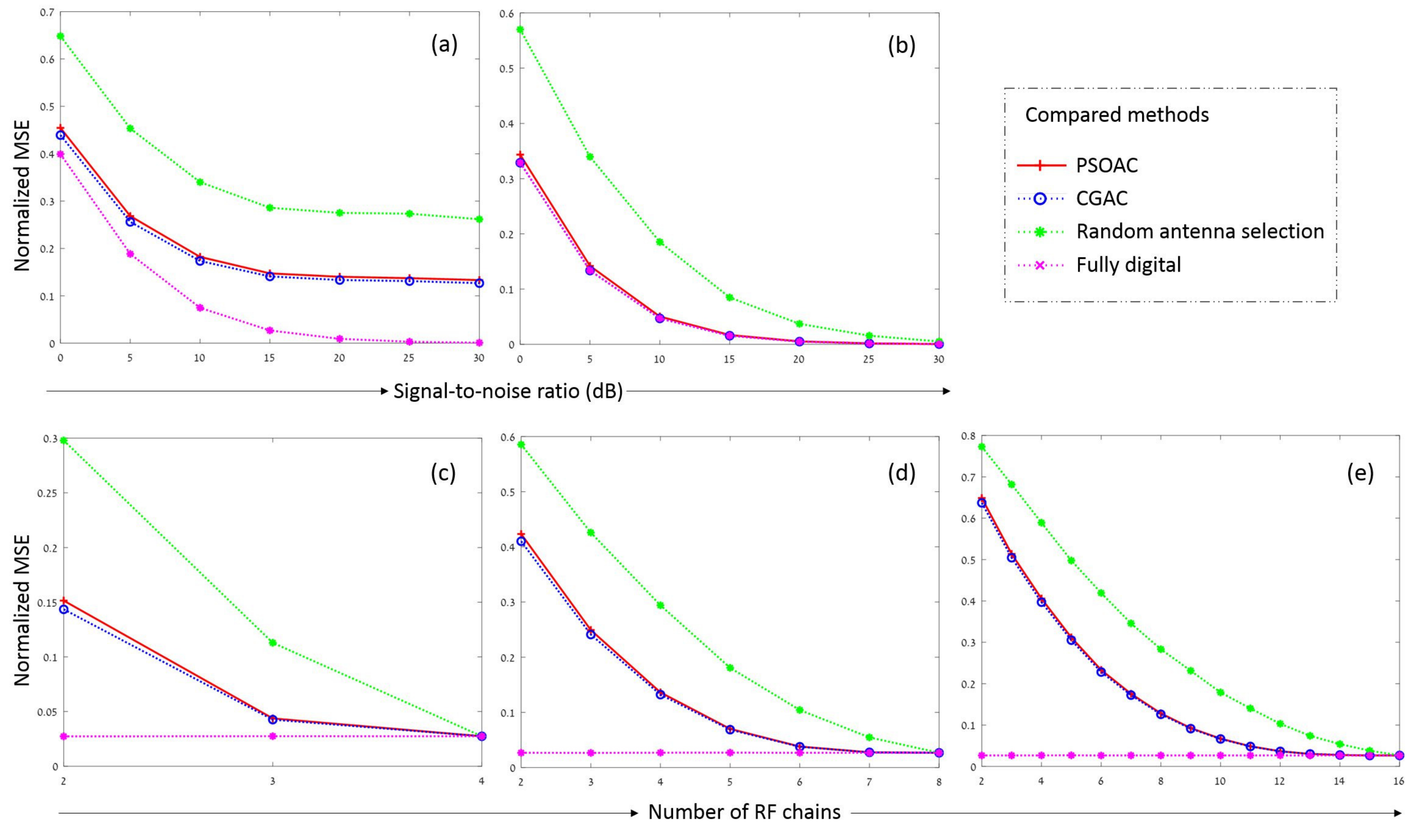}
	\caption{The normalized MSE curves for analog combiners versus SNR with (a) the regular rank case and (b) the best rank case of the receive side correlation matrix. The normalized MSE curves at \ac{snr} of $15$ dB and the regular rank case for the analog combiners with (c) 4 antennas, (d) 8 antennas and (e) 16 antennas.}
	\label{Results}
\end{figure*}

\section{Experiment Results}
\label{sec:ER}
	In the experimental study, we evaluate the channel estimation performance when the observed channel output is combined using our analog combining hardware, configured as both a \acs{cgac} as well as a \acs{psoac}. While we focus here on \ac{mse} channel estimation, the proposed analog combiner can be used to evaluate various other measures, such as bit error rate, by using different processing of the combiner output, e.g., symbol detection, in the digital domain.
	We implement \acp{cgac}, designed via \eqref{optimalW} by setting $\mathbf{V} = \mathbf{D} = \mathbf{I}_{N_{rf}}$, and \acp{psoac}, configured using Algorithm \ref{alg:algorithm}. The channel estimation accuracy is compared to that achievable when using a random antenna selection, which randomly selects $N_{rf}$ out of $N_{bs}$ antennas, and also to the traditional fully-digital setup, where each antenna is connected to a dedicated RF chain. 
	
	In all the presented experiments,  the number of \acp{ut} is fixed to $K = 3$, and the pilot sequence length is $\tau = K = 3$. 
	Following \cite{Ioushua2018Pilot}, we fix the transmit side correlation matrix $\mathbf{P}$ to the $K \times K$ identity matrix. The receive side correlation matrix $\mathbf{Q}$ is generated as follows: for a fixed $N_Q > 0$, we randomize an $N_{bs} \times N_Q$ proper-complex Gaussian matrix with {i.i.d.} zero-mean unit variance entries, denoted $\tilde{\mathbf{Q}}$, and set $\mathbf{Q} = \tilde{\mathbf{Q}}\tilde{\mathbf{Q}}^*$. We consider two settings of $N_Q$: $N_{rf} < N_Q \le N_{bs}$, referred to as the {\em regular setting}, and $N_Q = N_{rf}$, referred to as the {\em best setting}. We generate 1000 independent realizations of $\mathbf{Q}$. It is noted that the general clustered channel model \cite{Xu2002Spatial, Alk2015Limited} is a special case of our Kronecker channel model used here \cite{Ioushua2019Hybrid}.
	The performance metric is the normalized MSE $\frac{\| \mathbf{h} - \hat{\mathbf{h}} \|^2}{\| \mathbf{h} \|^2}$
	%, computed by averaging $ \frac{\E\{\| \mathbf{h} - \hat{\mathbf{h}} \|^2\}}{\E\{\| \mathbf{h} \|^2\}}$ 
	over the generated receiver correlation matrices. 
	For each Monte Carlo simulation, new realizations of the channel matrix, noise vector, and pilots matrix are generated. Such noise vector and channel matrix obey the model detailed in Section \ref{sec:SA}. The pilots matrices are obtained from $K$ eigenvectors of a $\tau \times \tau$ random matrix randomized from a proper-complex Gaussian distribution with {i.i.d.} zero-mean unit variance entries.

%-----------------------------------
%	Training Size Analysis
%-----------------------------------
\subsection{Normalized MSE versus SNR}
\label{subsec:SNR} 
	We first test the normalized \ac{mse} in estimating the channel versus \ac{snr} in the range $[0,30]$ dB, for an analog combiner with $N_{bs} = 8$ antennas and $N_{rf} = 4$ RF chains. 
	Note that this analog combiner model is obtained using the virtual channel extension approach detailed in Section \ref{sec:SA}. 
	The results are depicted in Figs. \ref{Results}(a)-\ref{Results}(b) for high rank (regular) receive correlation and for low rank (best) receive correlation, respectively. 
	Observing Figs. 6(a)-6(b), we note that using the proposed hardware prototype, the \acs{psoac} configured via Algorithm 1, which represents a practical family of analog combiners, achieves channel estimation accuracy within a very small gap compared to the costly \acs{cgac} architecture. 
    Furthermore, both \acs{psoac} and \acs{cgac} architectures notably outperform the random antenna selection approach. It is noted however that random antenna selection is simpler to implement compared to analog-combiner based systems, which require additional hardware. As expected, the fully-digital architecture achieves the lowest \acs{mse} performance, as it has access to the observed channel output without dimensionality reduction.

	For the high rank (regular) receiver correlation case, it is seen in Fig. \ref{Results}(b) that all approaches which implement analog combining meet an error floor at high \acp{snr} (above $20$ dB). This error floor is due to the model mismatch induced by the dimensionality reduction, which becomes the main performance bottleneck at this \ac{snr} regime. 
	However, when the rank of the receive correlation matrix is not larger than the number of RF chains, the optimal \ac{mse} performance observed using fully-digital receivers is also achievable using our analog combiner prototype, in line with the theoretical results of \cite{Ioushua2019Hybrid}.

%-----------------------------------
%	Error Rate Comparison
%-----------------------------------
\subsection{Normalized MSE versus Number of RF Chains}
\label{subsec:NRFC} 
	Next, we numerically evaluate the normalized \ac{mse} performance achievable using our hardware analog combiner prototype for different numbers of RF chains. Here, we let the number of RF chains $N_{rf}$ vary in the range of $[2, N_{bs}]$, for a fixed \ac{snr} level of 15 dB, and for a high rank (regular) receive correlation setup. The results of this experiment are depicted in Figs. \ref{Results}(c), \ref{Results}(d) and \ref{Results}(e), for $N_{bs} = \{4, 8, 16\}$, respectively.  Observing Figs. \ref{Results}(c)-\ref{Results}(e),  we note that for all considered scenarios the normalized MSEs of our hybrid architectures approach the normalized MSEs of the fully-digital receiver as the number of RF chains increases. This result settles with the fact that for $N_{rf} = N_{bs}$ these networks implement a fully-digital receiver. When the reduction rate $\frac{N_{rf}}{N_{bs}}$ is above  $62.5\%$, the performance gaps between the proposed analog combiners and the fully-digital receiver become negligible.  
	This behavior is due to the numerical observation that most simulated channels can be reliably characterized by roughly $5/8$ of their eigenmodes, which are reliably recovered and restored using our analog combiner hardware prototype. 	
	Moreover, among the hybrid receivers, the \ac{psoac} achieves normalized MSE values which are only slightly higher than that of the costly \ac{cgac}, emphasizing the benefits of its design via Algorithm \ref{alg:algorithm}. Finally, both the \ac{cgac} as well as the \ac{psoac}  have lower normalized MSEs than the random antenna selection. 
	
	These results demonstrate the ability of our proposed configurable analog combiner hardware to efficiently implement  desirable RF chain reduction while inducing minimal performance degradation on the overall communication system.

%----------------------------------------------------------------------------------------
%	CONCLUSIONS
%----------------------------------------------------------------------------------------
\section{Conclusions}
\label{sec:Conclusions}
	In this work, we presented a hardware prototype of a \ac{mimo} receiver with RF chain reduction via configurable analog combining. Our proposed prototype consists of a specially designed combiner board as well as a dedicated experimental setup which allows to test and adjust the analog combiner weights. We configure the analog combiner to optimize the channel estimation accuracy in \ac{mimo} systems by proposing an algorithm which improves upon state-of-the art design methods. Using our hardware prototype we were able to achieve \ac{mimo} channel estimation accuracy which is comparable to that achievable using costly fully-digital receivers.

\appendices
\section{Proof of Theorem 1}
\label{theorem1}
	Let $\mathbf{W} = \mathbf{V} \mathbf{\Lambda} \mathbf{U}^{*}$ be the singular value decomposition of $\mathbf{W}$, where $\mathbf{V} \in \mathbb{C}^{N_{rf} \times N_{rf}}$ and $\mathbf{U} \in \mathbb{C}^{N_{bs} \times N_{bs}}$  are its left and right singular matrices, respectively, and $\mathbf{\Lambda} \in \mathbb{C}^{N_{rf} \times N_{bs}}$ is  its  singular values diagonal matrix. 
	To prove the theorem, we first note that the objective \eqref{maximization} is invariant to the choice of the unitary $\mathbf{V} \in \mySet{U}$, and thus 
	\begin{flalign}
	\nonumber
	f( \mathbf{W} )&= f(\mathbf{\Lambda} \mathbf{U}^{*}) \Big.\\
	\nonumber
	&= tr \Big( (\mathbf{P} \otimes \mathbf{Q}) (\mathbf{S}^{*} \otimes \mathbf{U}\mathbf{\Lambda}^{*}) \Big.\\
	\nonumber
	&\quad \Big. \times\left[ (\mathbf{S} \otimes \mathbf{\Lambda}\mathbf{U}^{*}) ((\mathbf{P} \otimes \mathbf{Q}) + p_{n}\mathbf{I}_{Kn_{bs}}) (\mathbf{S}^{*} \otimes \mathbf{U}\mathbf{\Lambda}^{*}) \right]^{-1} \Big.\\
	&\qquad \qquad \qquad \qquad \Big. \times(\mathbf{S} \otimes \mathbf{\Lambda}\mathbf{U}^{*}) (\mathbf{P} \otimes \mathbf{Q})^{*} \Big). 
	\label{problemAppdix1}
	\end{flalign}	
	Next, we write $\mathbf{\Lambda} \mathbf{U}^{*} = \tilde{\mathbf{\Lambda}} \tilde{\mathbf{U}}^{*}$ where $\tilde{\mathbf{\Lambda}} \in \mySet{D}$ and  $\tilde{\mathbf{U}}\in \mathbb{C}^{N_{bs} \times N_{rf}}$  are the first $N_{rf}$ columns of $\mathbf{\Lambda}$ and $\mathbf{U}$, respectively. Using this formulation, it is noted that \eqref{problemAppdix1} is invariant to the setting of $\tilde{\mathbf{\Lambda}}$, and can be written as 
	\begin{flalign}
	\nonumber
	f( \mathbf{W} ) &=	f( \tilde{\mathbf{U}}^*)\Big.\\
	\nonumber
	&= tr \Big( (\mathbf{P} \otimes \mathbf{Q}) (\mathbf{S}^{*} \otimes \tilde{\mathbf{U}}) \Big.\\
	\nonumber
	&\quad \quad \Big. \times\left[(\mathbf{S} \otimes \tilde{\mathbf{U}}^{*}) ((\mathbf{P} \otimes \mathbf{Q}) + p_{n}\mathbf{I}_{Kn_{bs}}) (\mathbf{S}^{*} \otimes \tilde{\mathbf{U}}) \right]^{-1} \Big.\\
	\nonumber
	&\qquad \qquad \qquad \qquad \Big. \times(\mathbf{S} \otimes \tilde{\mathbf{U}}^{*}) (\mathbf{P} \otimes \mathbf{Q})^{*} \Big) \\
	\label{eqn:Problem2}
	&\!\overset{(a)}{=}\! K \cdot tr \!\left( \tilde{\mathbf{U}}^{*} \alpha^{2} \mathbf{Q}^{2} \tilde{\mathbf{U}} \left[ \tilde{\mathbf{U}}^{*} \left( \alpha \mathbf{Q} \!+\! p_{n} \mathbf{I}_{n_{bs}} \right) \tilde{\mathbf{U}} \right]^{\!-\!1}\! \right),
	\end{flalign}		
	where $(a)$ follows from the cyclic invariance property of the trace operator and by substituting  $\mathbf{P} = \alpha \mathbf{I}_{K}$ and $\mathbf{S}\mathbf{S}^* = \mathbf{I}_{\tau}$.
	The optimization in \eqref{eqn:Problem2} is then solved using $\tilde{\mathbf{U}} = {\mathbf{U}}^o$ by \cite[Prop. 2]{Ioushua2019Hybrid}, concluding the proof of the theorem. \qed

\section*{Acknowledgment}
	The authors are grateful to Eli Laks, Harel Moalem, Maxim Meltsin, Eli Shoshan, Amir Daichik, and Golan Robinsohn for their help and support in building this prototype.

\ifCLASSOPTIONcaptionsoff
\newpage
\fi

\bibliographystyle{IEEEtran}
\bibliography{IEEEabrv,references}

\end{document}